\def\preprint{preprint}
\def\review{review}

\def\myclass{\preprint}

\documentclass[\myclass]{elsarticle}

\usepackage{lineno}
\usepackage{hyperref}
\usepackage{multirow}
\usepackage{multicol}
\usepackage{graphicx}
\usepackage{longtable}
\usepackage{algpseudocode}
\usepackage{algorithm}
\usepackage{amsmath}
\usepackage{xcolor}
\usepackage{subcaption}
\usepackage[T1]{fontenc}
\usepackage{mathtools}
\usepackage{ifthen}

\ifthenelse{\equal{\myclass}{\review}}{\journal{Journal of Computational Science}}{\journal{Journal}}

\DeclarePairedDelimiter{\ceil}{\lceil}{\rceil}
\DeclareMathOperator*{\argmax}{arg\,max}

\algnewcommand\algorithmicforeach{\textbf{for each}}
\algdef{S}[FOR]{ForEachP}[1]{\algorithmicforeach\ #1\ \algorithmicdo~{\bf parallel}}
\ifthenelse{\equal{\myclass}{\review}}{\modulolinenumbers[5]}{\textwidth=426pt \hoffset=-40pt}
\newcommand{\sokol}{{\it sokol$_{\mathit{skew}}$}}
\newcommand{\sokolfull}{{\it sokol$_{\mathit{full}}$}}
\newcommand{\orel}{{\it lssOrel}}
\newcommand{\lastovka}{{\it xLastovka}}

\begin{document}

\begin{frontmatter}

\title{Parallel Self-Avoiding Walks for\\a Low-Autocorrelation Binary Sequences Problem}

\ifthenelse{\equal{\myclass}{\review}}{
    \author{Borko Bošković}
}
{
    \author{Borko Bošković~\href{https://orcid.org/0000-0002-7595-2845}{\includegraphics[width=0.25cm]{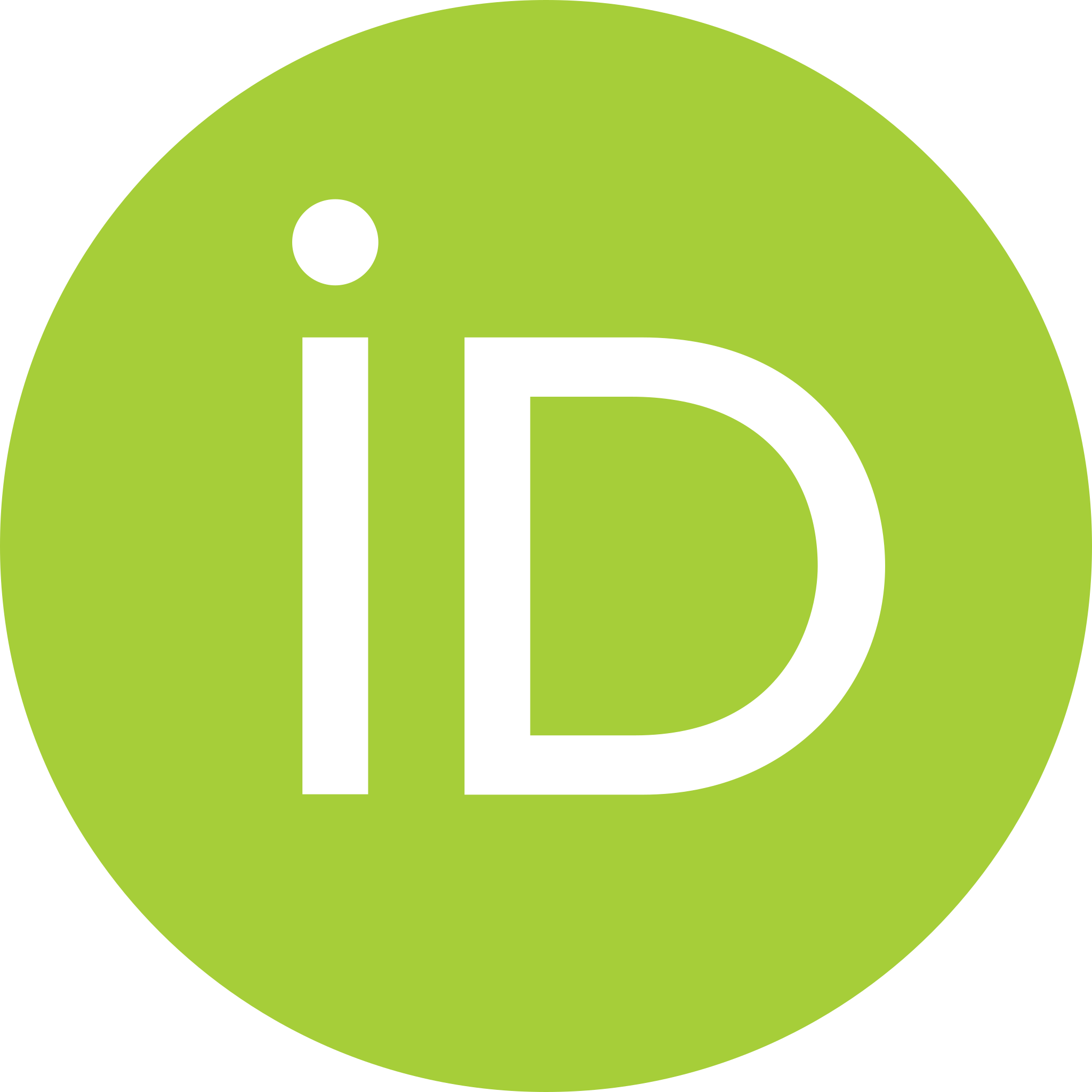}}}
}
\ead{borko.boskovic@um.si}

\ifthenelse{\equal{\myclass}{\review}}{
    \author{Jana Herzog}
}
{
    \author{Jana Herzog~\href{https://orcid.org/0000-0001-5555-878X}{\includegraphics[width=0.25cm]{orcid.png}}}}
\ead{jana.herzog1@um.si}

\ifthenelse{\equal{\myclass}{\review}}{
    \author{Janez Brest}
}
{
    \author{Janez Brest~\href{https://orcid.org/0000-0001-5864-3533}{\includegraphics[width=0.25cm]{orcid.png}}}
}
\ead{janez.brest@um.si}

\address{Faculty of Electrical Engineering and Computer Science,\\
University of Maribor, SI-2000 Maribor, Slovenia}

\begin{abstract}
A low-autocorrelation binary sequences problem with a high figure of merit factor represents a formidable
computational challenge. An efficient parallel computing algorithm is required to reach the new best-known solutions 
for this problem. Therefore, we developed the \sokol{} solver for the skew-symmetric search space. The developed solver
takes the advantage of parallel computing on graphics processing units. The solver organized the search process as a
sequence of parallel and contiguous self-avoiding walks and achieved a speedup factor of 387 compared with \orel{}, its
predecessor. The \sokol{} solver belongs to stochastic solvers and can not guarantee the optimality of solutions. To
mitigate this problem, we established the predictive model of stopping conditions according to the small instances
for which the optimal skew-symmetric solutions are known. With its help and 99\% probability, the \sokol{} solver found
all the known and seven new best-known skew-symmetric sequences for odd instances from $L=121$ to $L=223$. For
larger instances, the solver can not reach 99\% probability within our limitations, but it still found several new
best-known binary sequences. We also analyzed the trend of the best merit factor values, and it shows that as sequence
size increases, the value of the merit factor also increases, and this trend is flatter for larger instances.
\end{abstract}


\end{frontmatter}

\ifthenelse{\equal{\myclass}{\review}}{\linenumbers}{}

\section{Introduction}
General-Purpose computing on Graphics Processing Units (GPGPU) is used to reach better solutions in many fields,
such as in computer vision, graph processing, biomedical, financial analysis, and physical
simulation~\cite{Khairy19,Hwu11,Renc22,Luca21,Daniel21}. The reason for that is the computational power of Graphics
Processing Units (GPU). Therefore, these devices are used in High-Performance Computing (HPC) for solving real-world
problems. However, current GPU devices have some drawbacks and limitations that must be considered when designing a new
algorithm. The architecture, memory bandwidth, amount of memory, and memory speed are some of them. Despite these
limitations, we can implement a solver for reaching impressive results and new solutions.

In this paper, we proposed a parallel self-avoiding walk algorithm for solving a Low-Autocorrelation Binary Sequences
(LABS) problem that represents a formidable computational challenge. Here we have to locate binary
sequences $S_L^*$ of length $L$ with the highest value of merit factor $F$, as shown in Eq.~(\ref{eq:labs}). These
sequences are important in many  areas, such as communication engineering~\cite{Juan17,Zhao17,Zeng20}, statistical
mechanics~\cite{Borwein08,Bernasconi87,Leukhin15}, chemistry~\cite{Stadler61},
cryptography~\cite{Schmidt16,Packebusch16}, and mathematics~\cite{Gunther17,Jedwab13,Gunther18}.

\begin{align}
\label{eq:labs}
S_L &= \{s_1, s_2, ..., s_L\};~~~s_i \in \{+1,-1\} \nonumber \\
E(S_L) &= \sum_{k=1}^{L-1}C_k^2(S_L)~;~~~C_k(S_L) = \sum_{i=1}^{L-k} s_i \cdot s_{i+k} \\
S_L^{*} &=  \argmax_{S_L} F(S_L)~;~~~F(S_L) =\frac{L^2}{2\cdot E(S_L)}  \nonumber
\end{align}

Two solvers, \orel{} and \lastovka{}, were proposed in our previous works~\cite{Boskovic17,Brest18}. Both
are stochastic solvers that take the advantage of the mechanism for fast evaluation of neighbor solutions and are
limited to skew-symmetric sequences. The main difference between both solvers is in the memory usage that guides the
search process. The \orel{} uses a self-avoiding walk and a small hash table to prevent visiting the same pivot twice.
In contrast, \lastovka{} stores the large number of promising sequences into a priority queue and uses them for managing
the search process. Both solvers explore the skew-symmetric search space only. This means the dimension of the search
space is reduced by almost half ($D=\frac{L+1}{2}$), and, consequently, the solver can find (sub)-optimal sequences
significantly faster. The skew-symmetric search space is defined only for odd instances, as shown in
Eq.~(\ref{eq:skew}), and optimal skew-symmetric sequences may not be optimal for the entire search space.

\begin{align}
\label{eq:skew}
s_{D+i}=(-1)^i\cdot s_{D-i}~;~~~D = \frac{L+1}{2}~,~~~i = 1, 2, ..., D-1~,~~~L=5, 7, ...
\end{align}

The authors in~\cite{Kamil19} proposed a hybrid architecture that utilizes CPU and GPGPU processing units for the
agent-based memetic computational system. This system was used for solving the LABS problem on the entire search space
with the steepest descent and tabu search. The obtained maximum speedup was 46 with the tabu search for
$L=128$. The usefulness of GPGPU processing units for solving the LABS problem is evident from this result.
Motivated by this work, we proposed the \sokol{} solver for skew-symmetric sequences. It is implemented for the
NVIDIA GPU devices with the help of the CUDA (Compute Unified Device Architecture) toolkit. The \sokol{} solver uses
parallel self-avoiding walks and a recently published mechanism for efficient neighborhood
evaluation~\cite{Dimitrov21}. Both are not only computationally efficient but also require a small amount of memory.
They represent a suitable algorithm that prevents cycling, and a mechanism which can take the advantage of GPU devices
to speedup the energy calculation. Using them, \sokol{} obtained a speedup of 387 compared with the sequential CPU
solver and consequently found new best-known sequences.

The predictive model of stopping conditions for \sokol{} is established in this paper according to the instances
from $L = 71$ to $L = 119$. Using this model, we determined the stopping conditions needed by the solver to reach the
optimal skew-symmetric sequences with a probability of 99\%. With these stopping conditions, \sokol{} found all the
known and seven new best-known sequences for all odd instances from $L = 121$ to $L = 223$. The \sokol{} solver was also
used for solving larger instances up to $L=247$. In this case, it found all known and several new best-known
sequences. We also calculated the probability that the achieved sequences are optimal according to the predictive model.

Using all the best merit factor values, we established their trend model. It shows that the merit
factor values increase slightly as the sequence length increases, and this trend is flatter for larger sequence sizes.

According to the above mentioned, the main contributions of this paper are:
\begin{itemize}
 \item A parallel search with self-avoiding walks, designed for skew-symmetric
sequences, which takes the advantage of GPGPU devices.
 \item Skew-symmetric sequences that are optimal with a probability of 99\% according to the predictive model for
 instances from $L=121$ to $L=223$.
 \item The seventeen new best-known sequences for $L \le 247$.
 \item The trend of merit factors according to new best-known sequences.
\end{itemize}

The remainder of the paper is organized as follows. Related work is described in Section~\ref{related}. The proposed
\sokol{} solver is described in Section~\ref{sec:sokol}. The description of the experiments, analysis of the proposed
solver, and the obtained results are presented in Section~\ref{experiments}. Finally, the paper ends with a conclusion
in Section~\ref{conclusion}.

\section{Related work}
\label{related}
This section is divided into two parts since we address two fields in this paper. The first part is finding the
skew-symmetric binary sequences with the highest value of merit factor, and the second is the development of stochastic
solvers that take the advantage of GPGPUs.

\subsection{Binary sequences with the highest value of merit factor}
In~\cite{Golay82,Ukil15} it was shown that the upper bound of merit factor for the LABS problem could be 12.3248 or
10.23 as $L \rightarrow \infty$. With the help of construction methods, it is possible to form a sequence with the
merit factor value of 6.342061 or 6.4382~\cite{Jedwab13,Baden11}. The difference between the established upper bound and
the obtained merit factors by construction methods remains large. However, these methods are favorable for large
sequence sizes. In the survey~\cite{Jedwab05}, we can find the following challenge: ``Find a binary sequence S of length
$L$ > 13 for which $F \ge 10$.'' In~\cite{Ferguson2005}, the authors have  suspected that for $L > 250$ under
the skew-symmetric search space, we can expect $F \ge 10$. Currently, the optimal sequences are known for $L \le 66$,
and the skew-symmetric optimal sequences are known for $L \le 119$~\cite{Packebusch16}.

Two well-known search approaches exist to find (sub)-optimal sequences. The exact search allows finding the optimal
sequence that is time-consuming and inappropriate for long sequences. On the other hand, the stochastic search can
return a reasonable good sequence but it is not necessarily optimal. Therefore, the stochastic search is preferable for
larger instances.
In the literature, we can find the following stochastic approaches used for solving the LABS problem: local
search algorithm~\cite{Farnane18,Dimitrov21}, tabu search~\cite{Halim08}, memetic algorithm combined with tabu
search~\cite{Gallardo09}, self-avoiding walk technique~\cite{Boskovic17}, guiding the search process with a priority
queue~\cite{Brest18,Brest22}, etc. Currently best-known longer skew-symmetric sequences are published in~\cite{Brest22}.

\subsection{Stochastic solvers and graphics processing units}
A general metaheuristic framework for solving combinatorial optimization problems based on GPGPU is presented in
~\cite{Weyland13}. It enables efficient utilization of the GPU for the parallelization of an objective function and
takes the advantage of the computational power provided by modern GPUs. The usefulness of this framework was
demonstrated in the probabilistic traveling salesman problem with deadlines. The efficient implementation of the local
search for GPU was presented in~\cite{Schulz13}. The authors have used different profiling information and
best-known practice to identify bottlenecks and improve performance incrementally. They took into account efficient
solution evaluation on kernels, design of CPU–GPU interactions, and how efficiently they can evaluate large
neighborhoods whose fitness structures do not fit into the GPU's memory. A GPU implementation of Ant Colony Optimization
is presented in~\cite{Cecilia13}, and~\cite{Audrey13}. In both works, the algorithm was applied to the traveling
salesman problem. The parallel differential evolution algorithm for GPUs was designed in~\cite{Wang13}, and, in such a
way, the computational time was reduced  for high-dimensional problems. The advanced parallel cellular genetic algorithm
for GPU was developed in~\cite{Pinel13} for solving large instances of the Scheduling of Independent Tasks problem. The
research in~\cite{Pietron17} uses difficult black-box problems and popular metaheuristic algorithms implemented on
up-to-date parallel, multi-core, and many-core platforms, to show that the population-based techniques may have
benefited from employing dedicated hardware like a GPGPU or an FPGA. In the paper~\cite{Yasudo22}, the authors have
proposed a scalable implementation of the adaptive bulk search for solving quadratic unconstrained binary optimization
problems. This search algorithm combines a genetic algorithm in a CPU and local searches in multiple GPUs.

The paper~\cite{Kamil19} proposes a hybrid GPGPU architecture for the memetic evolutionary multi-agent systems to
improve the efficiency of computations. It is analyzed on the entire search space of the LABS problem, and the
speed-up factor 46 was obtained for $L = 128$ compared to the CPU solver. In~\cite{Dominik21}, the two new heuristics
for the whole search space of the LABS problem were developed and implemented on the GPGPU architectures. The authors
have shown that exploring the larger neighborhood improves the quality of the found sequences.

In contrast to the mentioned works on the LABS problem with GPGPU approaches, our approach is based on skew-symmetric
sequences. It uses self-avoiding walks with a recently published efficient neighborhood evaluation
mechanism~\cite{Dimitrov21}. Using these components, we designed an architecture where the amount of memory used by a
contiguous self-avoiding walk is minimized, the overhead related to synchronization between the CPU and GPGPU's memories
is minimized, and the possibility of cycling in the searching process is reduced. This way the implemented solver can
obtain a high speed or the number of sequence evaluations per second and find new best-known sequences.

\section{The \sokol{} solver}
\label{sec:sokol}

The way the solver explores the search space is crucial for its efficiency. In~\cite{Boskovic17} it was demonstrated
that cycling can happen within the search process of the skew-symmetric sequences. A sequence of contiguous
self-avoiding walks $\{\mathit{SAW}_{t_1}, \mathit{SAW}_{t_2},..., \mathit{SAW}_{t_N}\}$ is used in our solver to
mitigate this problem (see Eq.~(\ref{eq:saws})). The reason for that is the memory limitation of devices. It is
impossible to perform an unlimited self-avoiding walk and prevent cycling completely because we cannot store all
visited solutions in the memory for larger instances. Each contiguous self-avoiding walk $\mathit{SAW}_{t_i}$ of
length $n$ is a walk at the time $t_i$ on a lattice that does not visit the same node or pivot more than once, and
contains $n$ steps or pivots $\{P_1, P_2, ..., P_n\}$. In our case, each lattice node represents a sequence $S_j$ of
length $L$ and has $D$ neighbors. Each neighbor ($S_{k}$) differs from its pivot in only one sequence component. The
first pivot ($P_1$) is selected randomly, while the following pivot is the best neighbor that has not been visited
before. With the hash table, we can store the previous pivots and, in an efficient way, prevent cycling within the
search process of one contiguous self-avoiding walk.

\begin{eqnarray}
\label{eq:saws}
\mathit{SAWs} &=& \{\mathit{SAW}_{t_1}, \mathit{SAW}_{t_2},..., \mathit{SAW}_{t_N}\} \nonumber \\
\mathit{SAW}_{t_i} &=& \{P_1, P_2, ..., P_n\} \nonumber \\
P_1 &=& \{s_{1,1}, s_{1,2}, ..., s_{1,D}\},~~ s_{1,i} = \texttt{rand}\{-1, +1\} \nonumber \\
N_{k} &=& \{S_{i}~~|~~d(P_{k},S_{i})=1,~~S_{i} \neq P_m\},   \nonumber \\
&i& = 1, 2, ..., D,~~k = 2,3, ..., n,~~m=1,2,...,k-1\nonumber \\
P_{k+1} &=& \arg\min_{S_{i}\in N_{k}} E(S_{i}) \\ \nonumber
\end{eqnarray}

\begin{figure}[t!]
\centering
\includegraphics[width=\textwidth]{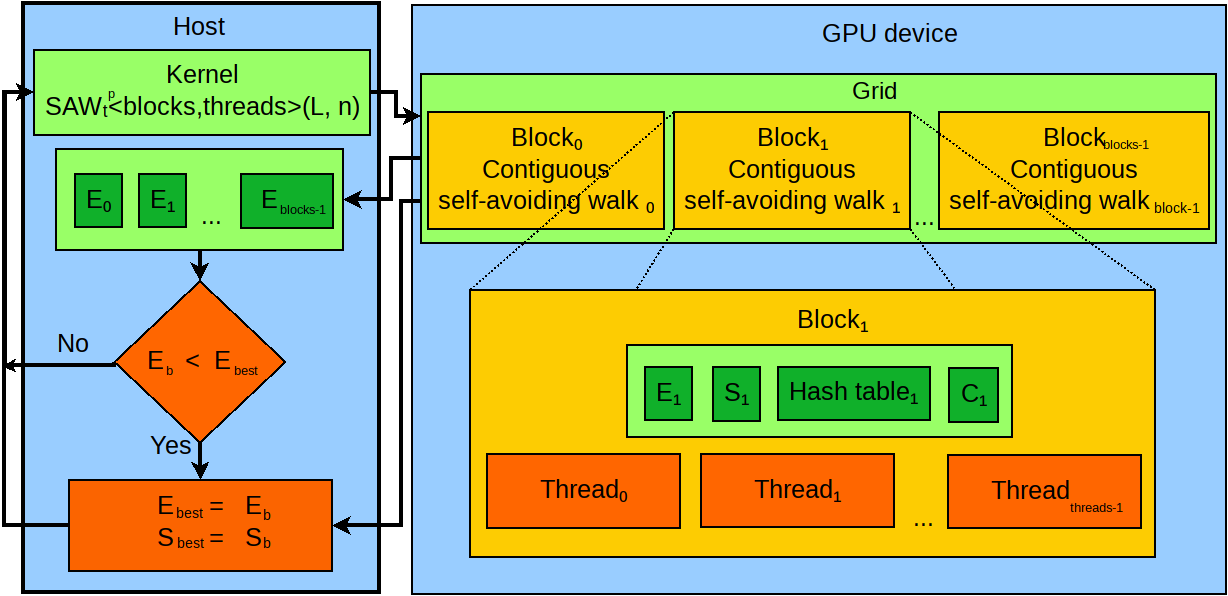}
\caption{Architecture of the \sokol{} solver.}
\label{fig:architecture}
\end{figure}

\begin{algorithm}[t!]
\caption{The \sokol{} algorithm}
\label{alg:sokol}
\begin{algorithmic}
\State{$\mathit{NFEs} = 0$}
\While{$t<t_{lmt}$}
    \State{\Call{$SAW^p_{t}${\it<<blocks, threads>>}}{$L,n$}}
    \State{$\mathit{NFEs} = \mathit{NFEs} + \mathit{blocks}*\mathit{n}*(D-1)$}
    \State{\{$E_0, E_1, ..., E_{blocks-1}$\} = \Call{TransferEnergiesToHost}{$blocks$}}
        \For{$b=0;~~b<blocks;~~b\text{++}$}
            \If{$E_b < E_{best}$}
                \State{$E_{best} = E_b$}
                \State{$S_{best} = \Call{TransferSequenceToHost}{b}$}
            \EndIf
        \EndFor
\EndWhile
\State{\Return{\{$E_{best}, S_{best}$\}}}
\end{algorithmic}
\end{algorithm}

\begin{eqnarray}
\label{eq:psaw}
\mathit{PSAWs} &=& \{\mathit{SAW}^p_{t_1}, \mathit{SAW}^p_{t_2},..., \mathit{SAW}^p_{t_N}\} \nonumber \\
\mathit{SAW}^p_{t_i} &=& \{\mathit{SAW}_{t_i}, \mathit{SAW}_{t_i}, ..., \mathit{SAW}_{t_i}\}  \\ \nonumber
\end{eqnarray}

Our solver organized the search process into a sequence of parallel searches $\{\mathit{SAW}^p_{t_1},
\mathit{SAW}^p_{t_2},...,$ $ \mathit{SAW}^p_{t_N}\}$. Each search performs parallel contiguous self-avoiding walks
$\{\mathit{SAW}_{t_i}, \mathit{SAW}_{t_i}, ..., \mathit{SAW}_{t_i}\}$ at the same time $t_i$, as shown in
Eq.~(\ref{eq:psaw}). All contiguous self-avoiding walks are independent. That means there is no need for
synchronization, and efficient parallelization is possible with GPGPU devices. However, the best-located sequence
and its energy value ($S_{best}$ and $E_{best}$) should be updated during the optimization process, as shown in
Algorithm \ref{alg:sokol} and Fig.~\ref{fig:architecture}. Parallel self-avoiding walks $SAW^p_{t}$ are performed in
each iteration. In the GPU device, each contiguous self-avoiding walk with the walk length $n$ is performed in one
block. Each block has several threads with shared memory. With these threads, each contiguous self-avoiding walk is
performed in parallel, as shown in Algorithm~\ref{alg:psaw}. Note that all functions in this algorithm are executed in
parallel and are responsible for the following:

\begin{itemize}
    \item generating the first pivot randomly,
    \item calculating the values of sidelobe array $C =\{ \hat{C}_0(S_L), \hat{C}_1(S_L), ..., \hat{C}_{L-1}(S_L) \}$
        where \\
        $\hat{C}_{L-k-1}(S_L) = C_k(S_L)$ for $k \in \{0, 1, ..., L-1\} $,
    \item calculating the differences $\Delta_{step}$ between the pivot $E_{step}$ and neighbors energy values,
    \item searching for the best-unvisited neighbor,
    \item updating the values of the sidelobe array and energy value for the next pivot $P_{step}$.
\end{itemize}

As already mentioned, each contiguous self-avoiding walk is performed in parallel with the specific number of
threads. This number depends on the instance size. It is set with Eq.~(\ref{eq:threads}) and this equation allows an
efficient parallel reduction mechanism in the following functions: \texttt{CalculateEnergy} and
\texttt{BestUnvisitedNeighbor}.
\begin{equation}
threads = 2^{(\ceil{\log_2(\frac{L+1}{2})})}
\label{eq:threads}
\end{equation}
The first function calculates the sum $\sum_{k=1}^{L-1}C_k^2(S_L)$, where $C_k^2(S_L)$
values are already calculated and stored in an array. The second function selects the best-unvisited neighbor. For
this purpose, we have used two arrays $\Delta_{step}$ and $\Delta_{index}$. The first array contains differences
between the pivot and neighbors' energy values. Exceptions are only neighbors that were already visited. Their
differences are set to the maximum values, which prevent them from being selected as the next pivot. The second array
contains the corresponding neighbor's indexes. Both are used in the reduction process, which returns the index of the
best-unvisited neighbor or next pivot.

\begin{algorithm}[t!]
\caption{$SAW^p_{t}${\it<<blocks, threads>>}($L,n$)}
\label{alg:psaw}
\begin{algorithmic}
\ForEachP{$block$}
\State{$P_1 = S_{block} = \Call{CreateRandomSequence<<{\it threads}>>}{L}$}
\State{$C_1 = \Call{CalculateSidelobeArray<<{\it threads}>>}{P_1}$}
\State{$E_{step} = E_{block} = \Call{CalculateEnergy<<{\it threads}>>}{C_1}$}
\State{$HashTable \leftarrow key(P_1)$}
\For{$step=1;~~step \le n;~~step\text{++}$}
    \State{$\Delta_{step} = \Call{CalculateNeighborhoodDelta<<{\it threads}>>}{P_{step},E_{step},C_{step}}$}
    \State{$P_{step+1} = \Call{BestUnvisitedNeighbor<<{\it threads}>>}{P_{step},\Delta_{step},HashTable}$}
    \State{$HashTable \leftarrow key(P_{step+1})$}
    \State{$\{E_{step+1}, C_{step+1}\} = \Call{UpdateData<<{\it
threads}>>}{P_{step+1},C_{step},E_{step},\Delta_{step}}$}
    \If{$E_{step+1} < E_{block}$}
        \State{$E_{block} = E_{step+1}$}
        \State{$S_{block} = P_{step+1}$}
    \EndIf
\EndFor
\EndFor
\end{algorithmic}
\end{algorithm}

When the parallel contiguous self-avoiding walks are finished, the best energy values from each block $E_{b}$ are
transferred from the GPU device to the host and compared with the best energy value ($E_{best}$). If a new best energy
value is found, the corresponding sequence $S_{block}$ is also transferred to the host, and the best sequence
($S_{best}$) is updated. This update is frequent only at the beginning of the optimization process, and only energy
values are often transferred to the host in each iteration. The overhead between the CPU and GPU memory is minimized
this way.

Each contiguous self-avoiding walk requires a specific data structures, the hash table for preventing cycles, and
a sidelobe array for efficient neighborhood evaluation, which contains only $L$ integers. From this, we can conclude
that a small amount of memory is required for one contiguous self-avoiding walk, and a large number of them can be
performed efficiently in parallel on GPU devices.

\section{Experiments}
\label{experiments}
The purpose of the experiments was to analyze the implemented \sokol{} solver. The solver was implemented with the C++
programming language using GNU C++ compiler 9.4 and the CUDA Toolkit 11.2.152.  The personal computer with an AMD Ryzen 
5 5600X processor was used for testing the \orel{} solver, and the NVIDIA A100-SXM4-40GB devices were used in the grid
environment VEGA\footnote{Vega is Slovenia’s petascale supercomputer - \url{https://doc.vega.izum.si/}.} for testing 
the \sokol{} solver.

\begin{figure}[t]
    \centering
    \includegraphics[width=0.49\textwidth]{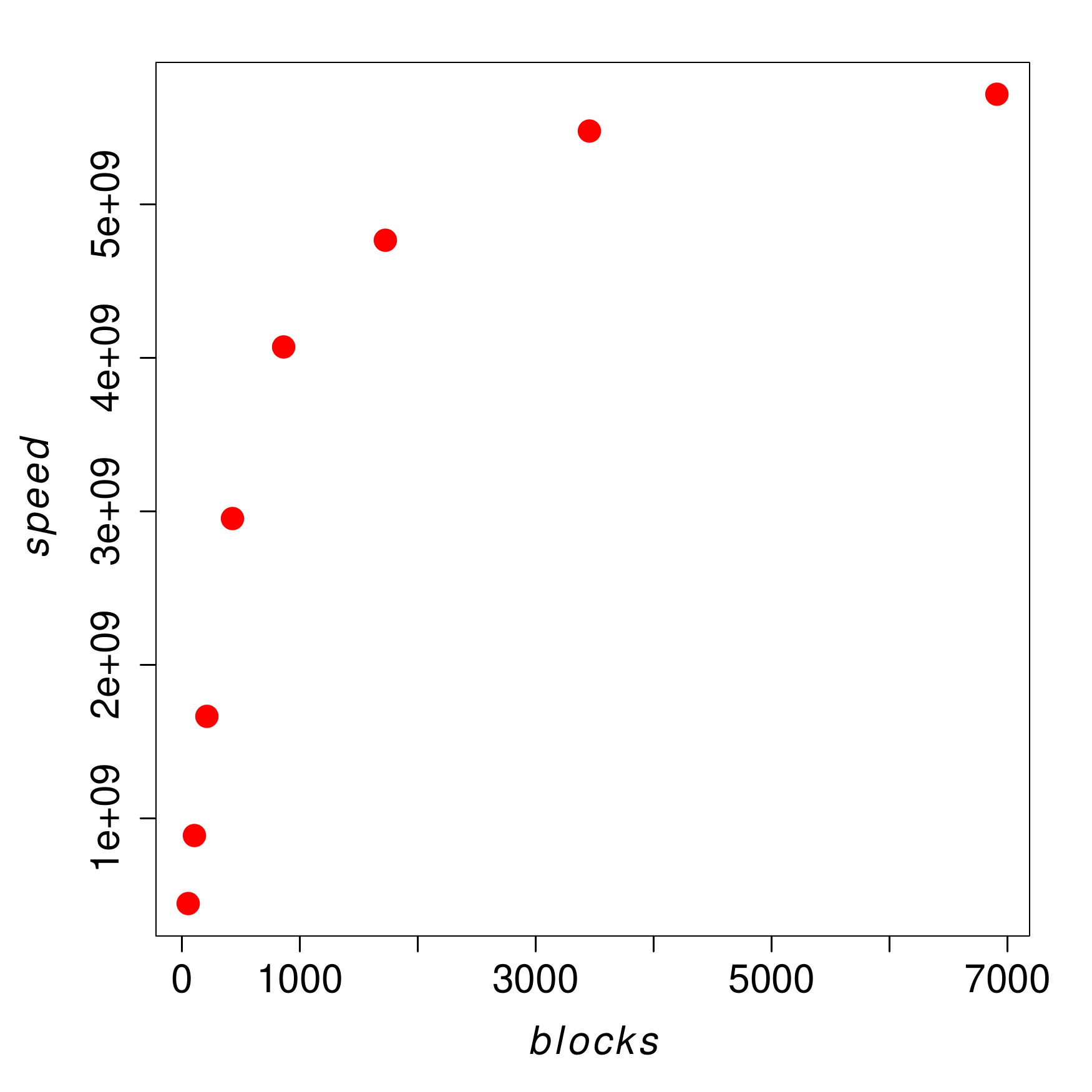}
    \caption{The ratio between variable $speed$ and $blocks \in \{54, 108, ...,
6912\}$ for $L=245$.}
    \label{fig:speedB}
\end{figure}

\subsection{Analysis of the \sokol{} solver}
In this section, we will analyze the efficiency of the \sokol{} solver. It has three control parameters $n$, $threads$,
and $blocks$. The first parameter defines the walk length of the contiguous self-avoiding walk. It is set to $n=8 \cdot
D$ according to~\cite{Boskovic17}. The second parameter, $threads$, determines the number of threads used for one
contiguous self-avoiding walk executed in one block. As was mentioned in the previous section, this number is
determined according to the sequence length (see Eq.~(\ref{eq:threads})). The third parameter, $blocks$, specifies the
number of blocks executed on the GPU device. This parameter minimizes the overhead related to synchronization between
the CPU and GPGPU memory. It affects the speed of the solver, and it is hardware dependent. Our GPU device has 6912
CUDA cores. To use all GPU cores, the product between the number of blocks and threads must be a multiple of the number
of CUDA cores. In our case these values could be 54 and 128, respectively. Therefore, the solver was analyzed on
sequence length $L=245$ with the following number of blocks 54, 108, 216, 432, 864, 1728, 3456, 6912 and with 128
threads, as shown in Fig.~\ref{fig:speedB}. From the displayed results, we can see that, as the number of blocks
increased, the speed of the solver also increased. We can not perform an infinite number of blocks, because a GPU
device has a limited amount of memory. Therefore, in all the following experiments, we will use the following equation
to calculate the number of blocks: $blocks=\frac{6912}{threads}\cdot 128$.

\begin{figure}[t!]
    \begin{subfigure}{0.49\textwidth}
        \centering
        \includegraphics[width=\textwidth]{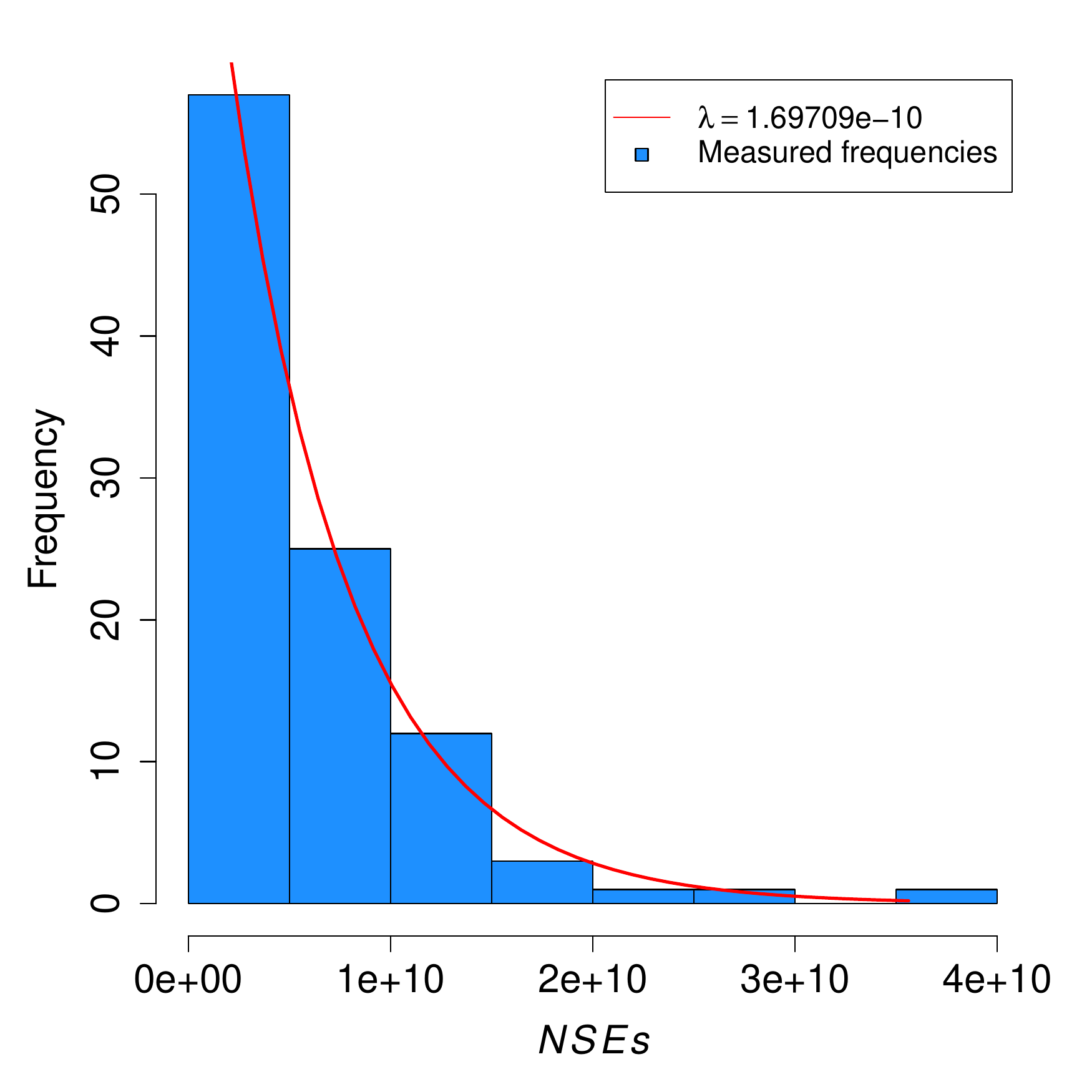}
        \caption{The distribution of $runtime$ for $L=117$.}
        \label{fig:distribution}
    \end{subfigure}
    \begin{subfigure}{0.49\textwidth}
        \centering
        \includegraphics[width=\textwidth]{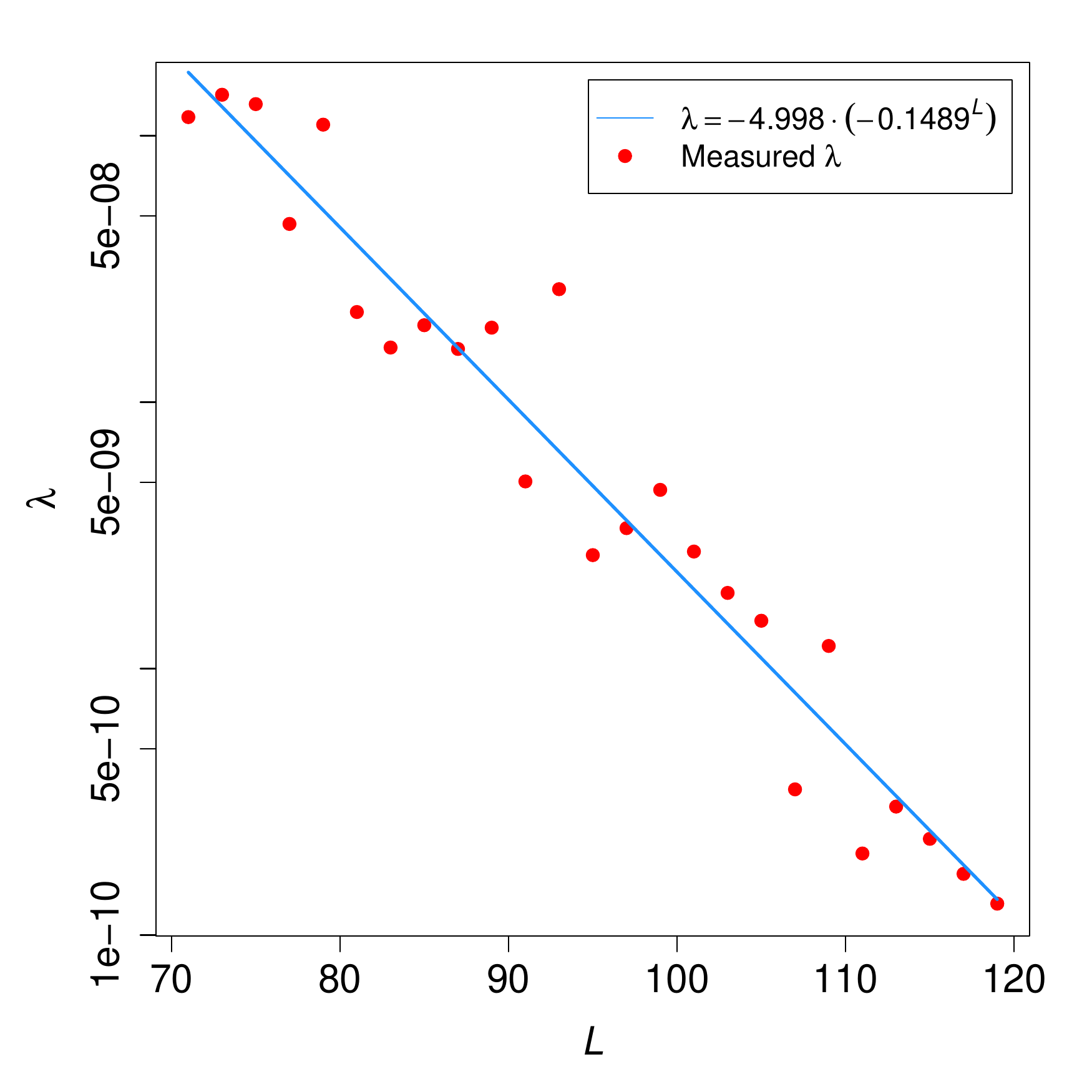}
        \caption{$\lambda$ for $L \in \{71, 73, ..., 119\}$.}
        \label{fig:lambda}
    \end{subfigure}
    \caption{The distribution of $\mathit{NSEs}$ that are needed to reach the optimal solution and the model
($R^2=0.9429$) of the exponential distribution parameter $\lambda$ according to the results for $L \in \{71, 73, ...,
119\}$.}
    \label{fig:nses}
\end{figure}

The main goal of our solver is to find optimal or sub-optimal solutions. Unfortunately, it belongs to stochastic
solvers, and cannot guarantee the optimality of solutions. However, we can establish the predictive model of stopping
conditions or the maximum number of solution evaluations ($\mathit{NSEs}_{lmt}$). According to this model, we can also
assume the optimality of reached solutions. For this purpose, the \sokol{} was analyzed on odd instances from L=71 to
L=119. These sequences were selected because the optimal solutions are known for them~\cite{Packebusch16}. For
determining $\mathit{NSEs}_{lmt}$, the target approach was used with the optimal solutions and 100 independent runs for
each instance size. With the Anderson-Darling test~\cite{Razali11} and visual identification, we found that the
variable $\mathit{NSEs}_{lmt}$ was distributed exponentially for all instances. For example, the histogram and
corresponding exponential distribution for $L=117$ are shown in Fig.~\ref{fig:distribution}. The $\lambda$ values of
exponential distributions for all the selected instances and their trend model ($\lambda=0.006753\cdot(0.8617^L)$) are
displayed in Fig.~\ref{fig:lambda}. This model will be used in the continuation of the paper for two purposes. The
first purpose was to determine the stopping condition $\mathit{NSEs}_{lmt}$ needed by the solver to reach the optimal
solution with a 99\% probability. For larger instances, the solver cannot obtain the determined  $\mathit{NSEs}_{lmt}$
due to $runtime_{lmt}=4$ days. Therefore, the second purpose was to calculate the probability of solutions' optimality
according to the predicted exponential distribution and obtained $\mathit{NSEs}$ values.

\subsection{The best-known sequences}
To demonstrate the efficiency of the proposed solver and locate new best-known sequences, we performed $N=100$
independent runs for $L \in \{121, 123, ..., 247\}$. The stopping conditions were $\mathit{NSEs}_{\mathit{lmt}}$
and $\mathit{runtime}_{\mathit{lmt}}$. $\mathit{NSEs}_{\mathit{lmt}}$ was defined for one run, and was set according
to the probability of 99\% with Eq.~(\ref{eq:stop}).
\begin{equation}
  \mathit{NSEs}_{lmt}(L) = \frac{\ln(1-P)}{-\lambda \cdot N} = \frac{\ln(1-0.99)}{-0.006753\cdot(0.8617^L)\cdot 100}
\label{eq:stop}
\end{equation}
The $\lambda$ model described in the previous section is used
in this equation. With the variable $N$ we also consider 100 independent runs. We can do this, because
contiguous self-avoiding walks within the search process are independent, and each of the 100 runs can be considered as
a single run. Each run was also limited with the $\mathit{runtime}_{\mathit{lmt}}=4$ days. This limitation was
determined by the VEGA environment. These values of the runtime and number of runs were also used in previous
works~\cite{Boskovic17,Brest22}. For instances where the solver cannot reach $\mathit{NSEs}_{\mathit{lmt}}$ within
$\mathit{runtime}_{\mathit{lmt}}=4$ days, we calculated the probability of reaching the optimal solution with
Eq.~(\ref{eq:probability}). The sum of the $\mathit{NSEs}$ for all independent runs was used in this equation because
we can consider $100$ runs as a single run.

\begin{equation}
 P(S_{best})=1-e^{-\lambda \cdot ( \sum_{i=1}^{N}\mathit{NSEs_i})}
 \label{eq:probability}
\end{equation}

With the described limitations, our solver reached the 17 new best-known sequences shown in Table~\ref{tab:seq}. These
sequences are presented by using a hexadecimal coding. Only first half of each sequence was coded, because the remaining
values are defined by skew-symmetry. Each hexadecimal digit is presented with a binary string: $0 = 0000, 1 = 0001, 2 =
0010, ..., F = 1111$.  Therefore, it is necessary to remove the leading 0 values to obtain the length of the binary
string $D$. To get the sequence values, each $0$ of the binary string should be converted to $+1$ and each $1$ to $-1$
or vice versa. In Table~\ref{tab:seq}, we can observe that the sequences obtained up to $L=223$ are optimal, with the
probability of 99\% according to the $\lambda$ model. For larger instances, the runs were limited to 4 days, which
means the corresponding probability is less than 99\%. For the largest instance size $L=247$, this probability was
reduced to only 13\%.

\begin{table}[t!]
\centering
\label{newsequences}
\caption{The new best-known skew-symmetric sequences obtained by the \sokol{} solver.}
\label{tab:seq}
\begin{tabular}{@{~}c@{~}|@{~}c@{~}|@{~}c@{~}|@{~}c@{~}|@{~}l@{~}}
$L$ & $E(S_{best})$ & $F(S_{best})$ & $P(S_{best})$ & $S_{best}$  \\
 \hline
 \hline
 171 & 1669 & 8.7600 & 99\% & \texttt{0x07F018C27F3C01849035B3}\\ \hline
 185 & 1932 & 8.8574 & 99\% & \texttt{0x0119ED2F78CF6800A4DE0623} \\ \hline
 193 & 2040 & 9.1296 & 99\% & \texttt{0x020C18D1A749035A04EFECC5A} \\ \hline
 197 & 2162 & 8.9752 & 99\% & \texttt{0x11556D25B59128BF09CDD2641} \\ \hline
 199 & 2187 & 9.0537 & 99\% & \texttt{0x0B09049607E02FB345D6C88E7} \\ \hline
 219 & 2605 & 9.2056 & 99\% & \texttt{0x0F1B163E62ACAA8F7814BF89231D} \\ \hline
 223 & 2727 & 9.1179 & 99\% & \texttt{0x03DC43EE6531A21CD95E148C084A} \\ \hline
 225 & 2768 & 9.1447 & 98\% & \texttt{0x06AF8A172B0EB88ADF54E5A74C629} \\ \hline
 229 & 2810 & 9.3311 & 87\% & \texttt{0x0F81FF03DFF1E7BCE6CB9B1517328} \\\hline
 231 & 2963 & 9.0046 & 78\% & \texttt{0x0240D99121A078037EFF306D34A2D} \\\hline
 235 & 2965 & 9.3128 & 57\% & \texttt{0x2D663B94D7EBFBD5B4884CA45ED23C} \\\hline
 237 & 3118 & 9.0072 & 46\% & \texttt{0x6D663B94D7EBFBD5B4884CA45ED23C} \\\hline
 239 & 3055 & 9.3488 & 37\% & \texttt{0xB64DB6017C0BAB48183C45C48C1A76} \\\hline
 241 & 3216 & 9.0300 & 29\% & \texttt{0x0B64DB6017C0BAB48183C45C48C1A76} \\\hline
 243 & 3233 & 9.1322 & 23\% & \texttt{0x2E7FC23843DADB804E1B3771FBE57E3} \\\hline
 245 & 3226 & 9.3033 & 17\% & \texttt{0x1C38F1EFD72180453AC7548DCFC5F19} \\\hline
 247 & 3259 & 9.3601 & 13\% & \texttt{0x3FF9FE03FE31FDEC1870F23887276E5} \\\hline
\end{tabular}
\end{table}

\begin{figure}[t!]
    \centering
    \begin{subfigure}{\textwidth}
        \centering
        \includegraphics[width=\textwidth]{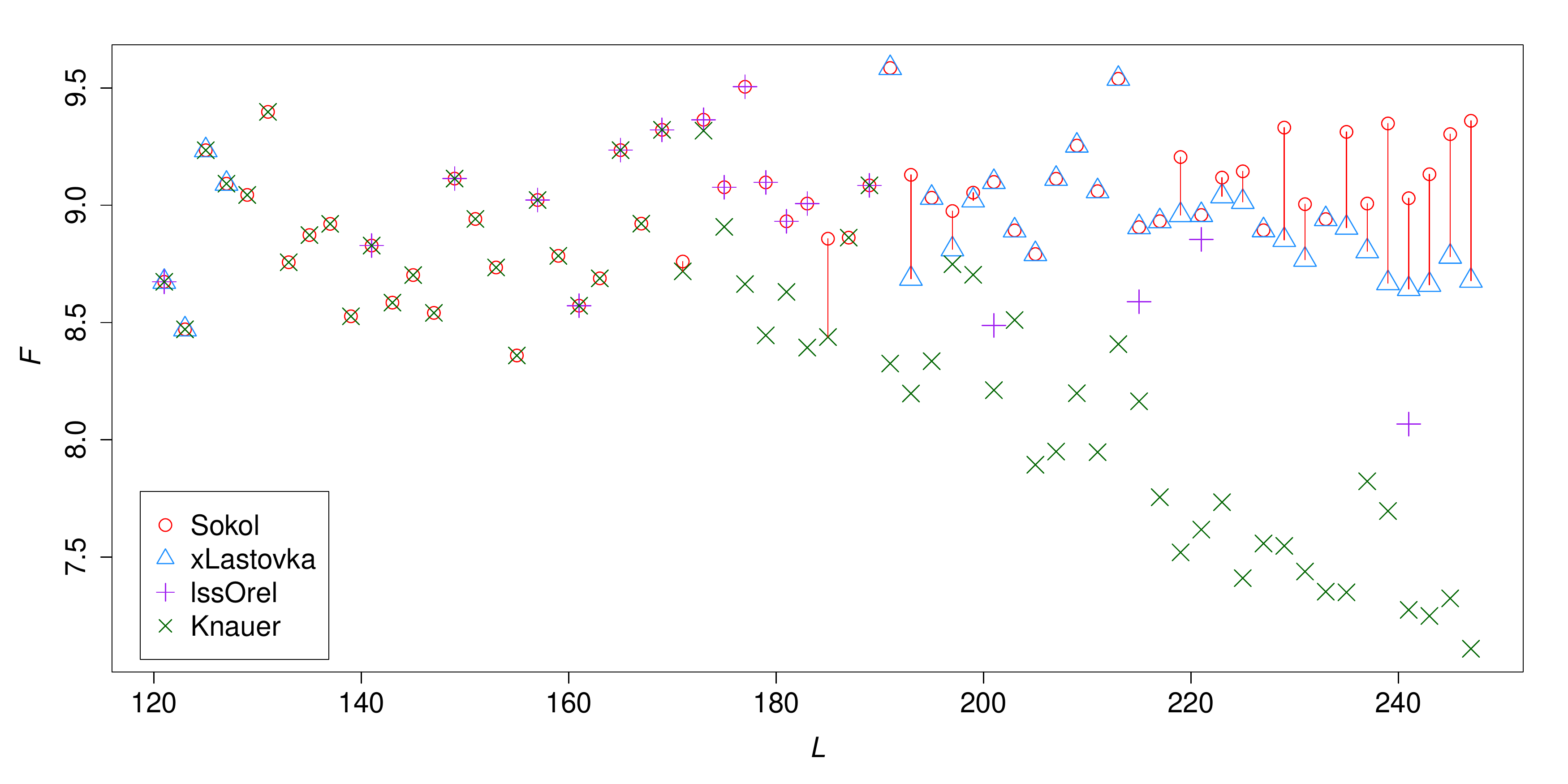}
        \caption{The best reached merit factors.}
        \label{fig:mf}
    \end{subfigure}
    
    \begin{subfigure}{\textwidth}
        \centering
        \includegraphics[width=\textwidth]{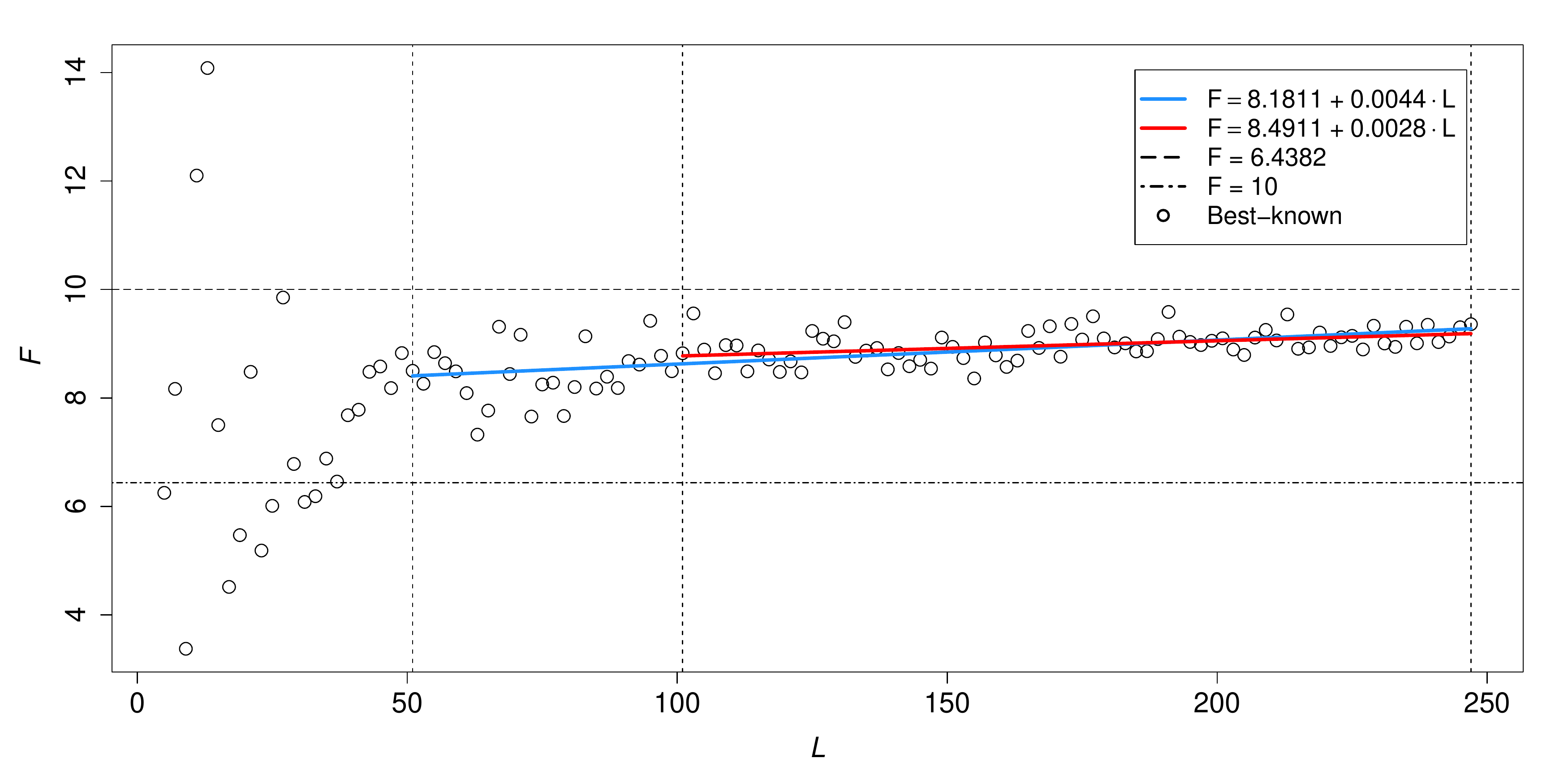}
        \caption{The best-known merit factors and their trend lines.}
        \label{fig:omf}
    \end{subfigure}
    \caption{The best merit factors.}
    \label{fig:mm}
\end{figure}

\begin{figure}[t!]
    \begin{subfigure}{\textwidth}
        \centering
        \includegraphics[width=\textwidth]{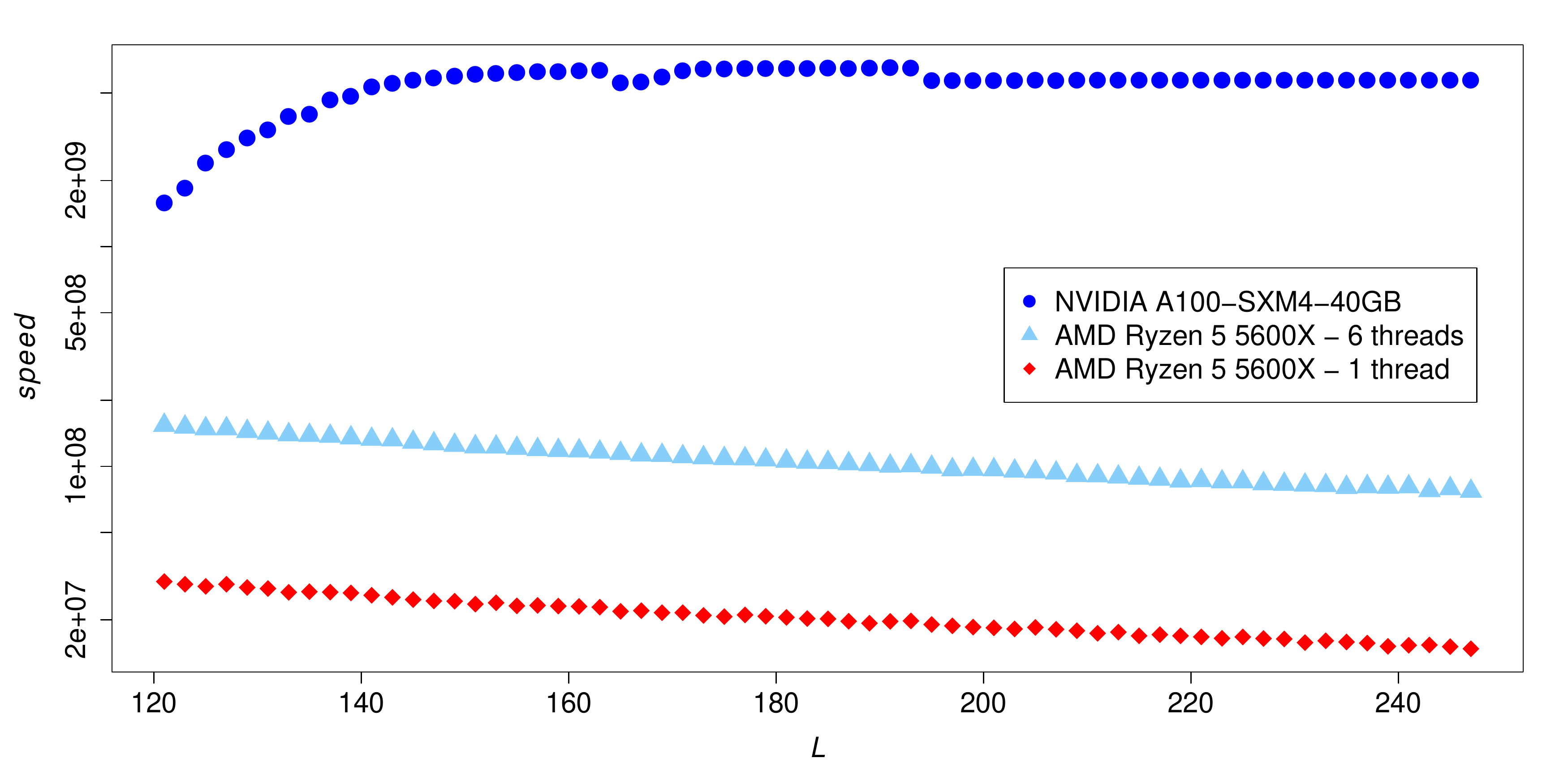}
        \caption{The $speed$ or the number of solution evaluations per second for $L \in \{121, 123, ..., 247\}$.}
        \label{fig:speedL}
    \end{subfigure}

    \begin{subfigure}{\textwidth}
        \centering
        \includegraphics[width=\textwidth]{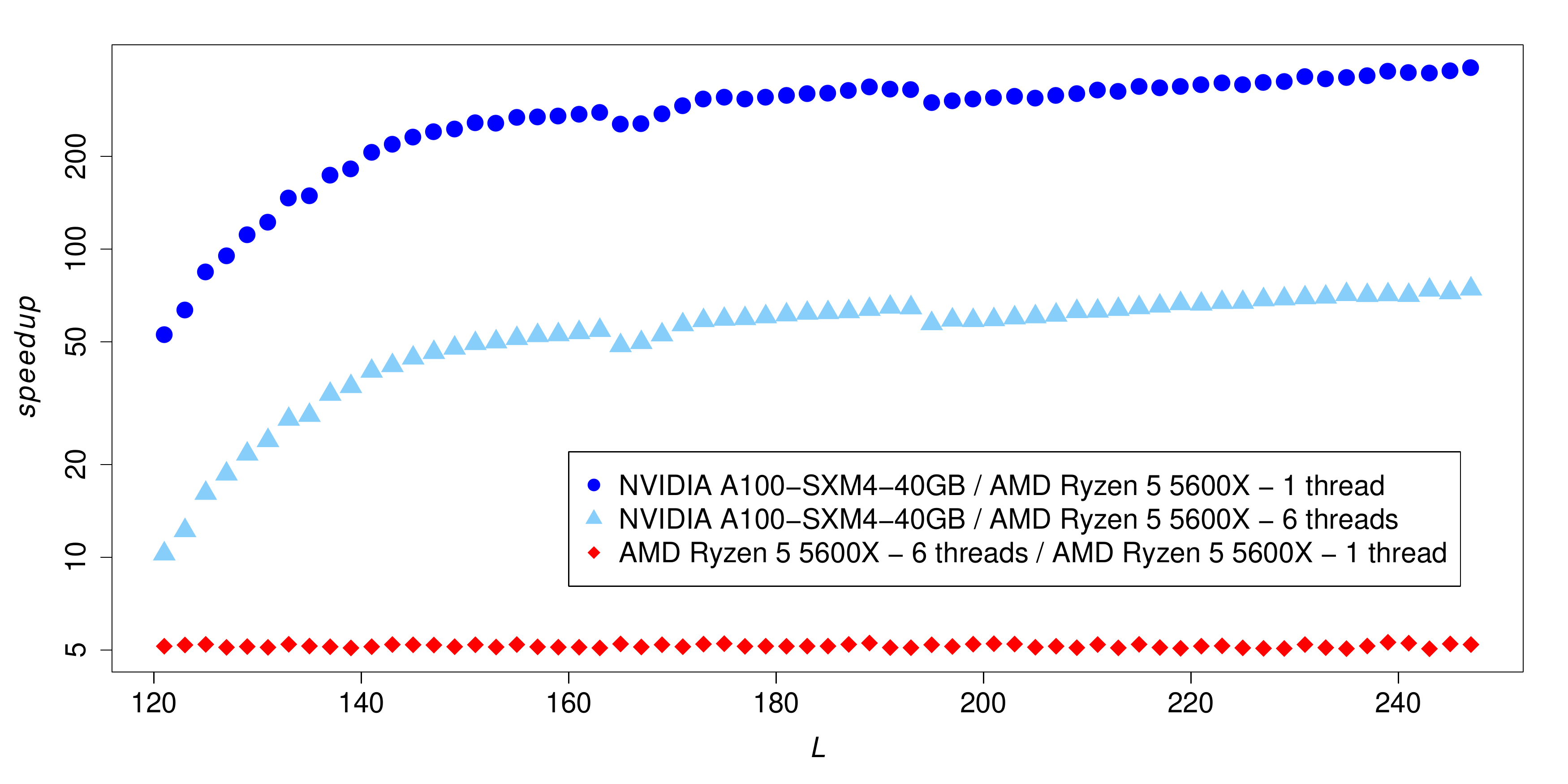}
        \caption{The $speedup$ or ratio between corresponding speeds for $L \in \{121, 123, ..., 247\}$.}
        \label{fig:speedup}
    \end{subfigure}
\label{speed}
\caption{The $speed$ ratio between the \sokol{} with 6912 GPU threads, parallel \orel{} with 6 CPU threads, and
sequential \orel{} solver.}
\label{fig:ss}
\end{figure}

The best-known merit factor values of our and related works~\cite{Knauer04,Boskovic17,Brest22}, are depicted in
Fig.~\ref{fig:mf}. As we can see, \sokol{} has found all the best-known sequences and 17 new best-known sequences.
Vertical red lines show improvements in these sequences. We can also observe that the significant improvements were
obtained the for the last seven instances. Additionally, merit factor values greater than 9 were obtained for
almost all new best-known sequences. In Fig.~\ref{fig:mf}, we can also observe the significant improvement reached in
the last two decades by comparing Knauer’s and our results. For example, the best merit factor value of Knauer's
result for $L=247$ was 7.1089, while our solver obtained a value of 9.3601. Fig.~\ref{fig:omf} shows the best-known
merit factors, including the merit factors' values found in this work, and two trend lines. The first line is fitted to
merit factor values from $L = 51$ to $L = 247$, while the second is fitted from $L = 101$ to $L = 247$. From the shown
lines, we can conclude that, as the sequence size increases, the value of the merit factor also increases. We can also
observe that the second fitted line is flatter, which means this increase of merit factor value for larger instances is
lower. However, the merit factors of the proposed \sokol{} solver are much better than the values of constructive 
methods (6.4382), but still below value 10, which is a challenge value for $L>13$. The largest merit factor value
found by \sokol{} was 9.3601 for $L=247$.

According to the reached new best-known solutions, it is evident that the proposed \sokol{} solver took the advantage of
parallel computing on GPU devices. It is also shown in Fig.~\ref{fig:mf} where we compare the results of the
\sokol{} solver and related works. Note that the same number of runs $N=100$ and $runtime_{lmt}=4$ days were used by
\orel{} and \lastovka{}. To demonstrate this advantage, Fig.~\ref{fig:speedL} shows the $speed$ or number of
sequence evaluations per second of the sequential \orel{} solver, parallel \orel{} solver with 6 CPU threads, and
\sokol{} solver with 6912 GPU threads. The obtained $speedup$ is shown in Fig.~\ref{fig:speedup}. The smallest
$speedup$ was achieved between the parallel and sequential \orel{}. The larger $speedup$ was obtained between the
\sokol{} and parallel \orel{}. In this case, the largest $speedup$ of 74 was obtained for $L=247$. The largest $speedup$
was obtained between the \sokol{} and sequential \orel{}. In this case, the $speedup$ of 387 was obtained for
$L=247$. From all the reported results, it is evident that the \sokol{} solver exploited the advantage of
parallel computing on GPU devices successfully and achieved outstanding results.

\section{Conclusions}
\label{conclusion}
This paper introduces a new stochastic \sokol{} solver for the low-autocorrelation binary sequences problem. It takes
the advantage of parallel computing on the graphics processing units. For this purpose, \sokol{} organizes the search
process as a series of parallel and contiguous self-avoiding walks on the skew-symmetric sequences. With these
sequences, the dimension of the search space is reduced  by almost half, and, consequently, the solver can find
(sub)-optimal sequences significantly faster. The skew-symmetric search space is defined for odd sequence lengths, and 
its optimal solutions may not be optimal for the entire search space. The contiguous self-avoiding walks reduce the 
possibility of cycling in the searching process and are independent of one another. The neighborhood evaluation
mechanism allows the solver to achieve a high speed in the search process and requires a small amount of memory. With
this mechanism, we designed an architecture that minimizes the overhead related to synchronization between the CPU and
GPU's memories. All the described components maximize the solver's speed or the number of solution evaluations per
second. This way, a bigger search space is examined, and a better sequence can be achieved at the same amount of
time.

The solver was analyzed to determine a good value of control parameters and stopping condition (the number of
solution evaluations) needed to reach an optimal solution. An exponential distribution was identified for
this stopping condition. Using the parameter $\lambda$ of the exponential distribution, we established the predictive
model for the stopping condition with a 99\% probability. With this model and the VEGA supercomputer, we applied the
proposed solver to all odd sequence lengths from 121 to 223. For these instances, the seven new best-know sequences were
found. For larger sequence sizes up to 247, the corresponding probability of optimality is calculated according to the
predictive model. In this case, the solver found ten new best-known sequences. The advantage of the proposed solver is
demonstrated by comparing the speed or number of solution evaluations per second between \sokol{} and its predecessor,
\orel{}. The \sokol{} solver with 6912 GPU threads obtained a speedup of 387 and 74 compared to sequential and parallel
\orel{} with 6 CPU threads, respectively.

The quality of sequences defines the merit factor. In this work, the maximum value of the merit factor was 9.3601 for
the sequence with length of 247. Unfortunately,  the challenge of finding a binary sequence of length greater than 13
for which the merit factor is greater or equal to 10, remains open. We also analyzed the trend of merit factor values.
This analysis shows that as the sequence size increases, the value of the merit factor also increases, and a flatter
trend is observed for larger instances.

In future work, we will try to improve the search algorithm or find a better search space. With the better search 
space, we will try to increase the probability of locating better solutions. We will also implement the \sokolfull{}
solver for the GPGPU devices that will examine the entire search space of the problem.

\section*{Acknowledgements}
This work was supported by the Slovenian Research Agency (Computer Systems, Methodologies, and Intelligent Services)
under Grant P2-0041.

\section*{CRediT authorship contribution statement}
{\bf Borko Bo\v{s}kovi\'{c}}: Conceptualization, Methodology, Software, Validation, Formal analysis, Investigation,
Writing - Original Draft, Visualization.
{\bf Janez Brest}, {\bf Jana Herzog}: Validation, Formal analysis, Investigation, Writing- Reviewing and Editing

\section*{Declaration of Competing Interest}
The authors declare that they have no known competing financial interests or personal relationships that could have
appeared to influence the work reported in this paper.

\bibliography{paper}

\end{document}